\documentclass[12pt,a4paper]{article}
\usepackage{amssymb,amsmath}
\usepackage[dvips]{lscape,graphicx}

\newcommand{\bc}{\begin{center}}
\newcommand{\ec}{\end{center}}
\newcommand{\bd}{\begin{displaymath}}
\newcommand{\ed}{\end{displaymath}}
\newcommand{\be}{\begin{equation}}
\newcommand{\ee}{\end{equation}}
\newcommand{\ba}{\begin{array}}
\newcommand{\ea}{\end{array}}
\newcommand{\bt}{\begin{tabular}}
\newcommand{\et}{\end{tabular}}

\newcommand{\ds}{\displaystyle}

\begin{document}

%\large

\title{Cosmological constant in SUGRA models and \\
the multiple point principle}

\author{C.Froggatt${}^{1}$, L.Laperashvili${}^{2}$, R.Nevzorov${}^{2}$,
H.B.Nielsen${}^{3}$\\[15mm] \itshape{${}^{1}$ Department of
Physics and Astronomy,}\\[3mm] \itshape{Glasgow University,
Glasgow, Scotland}\\[3mm] \itshape{${}^{2}$ Theory Department,
ITEP, Moscow, Russia}\\[3mm] \itshape{${}^{3}$ The Niels Bohr
Institute, Copenhagen, Denmark }}

\date{}

\maketitle

\begin{abstract}{
\noindent
The tiny order of magnitude of the cosmological constant is
sought to be explained in a model involving the following ingredients:
supersymmetry breaking in N=1 supergravity and the multiple point
principle. We demonstrate the viability of this scenario in the
minimal SUGRA model.}
\end{abstract}

\vspace{7cm} \footnoterule{\noindent${}^{1}$ E-mail:
c.froggatt@physics.gla.ac.uk\\ ${}^{2}$ E-mail:
laper@heron.itep.ru, nevzorov@heron.itep.ru\\ ${}^{3}$ E-mail:
hbech@alf.nbi.dk}

\newpage
\section{Introduction}

As is well known, cosmology yields strong arguments against the
Standard Model (SM). Although the SM describes perfectly the major
part of all experimental data measured in earth based experiments,
it does not provide any reliable candidate for dark matter.
Another puzzle of modern cosmology is a tiny density of energy
spread all over the Universe (the cosmological constant), which is
responsible for its acceleration. At first glance this
energy density ($\Lambda$) should be of the order of the Planck, or
possibly the electroweak, scale to the fourth power;
however a fit to the recent
data shows that $\Lambda \sim 10^{-123}M_{Pl}^4 \sim 10^{-55}
M_Z^4$ \cite{0}.
The smallness of the cosmological constant should be considered as
a fine-tuning problem, for which new theoretical ideas must be
employed to explain the enormous
cancellations between the contributions of different condensates
to the cosmological constant.

Unfortunately the cosmological constant problem can not be resolved in
any available generalization of the SM. An exact global supersymmetry
(SUSY) ensures zero value for the energy density at the minimum of the
potential of the scalar fields. But in the exact SUSY limit bosons and
fermions from one chiral multiplet get the same masses. Soft
supersymmetry breaking, which guarantees the absence of
superpartners of observable fermions in the $100\,\mbox{GeV}$
range, does not protect the cosmological constant from an
electroweak scale mass and the fine-tuning
problem is re-introduced.

In this article we propose the multiple point principle (MPP)
\cite{2} as a basic principle to explain the size of the
cosmological constant. MPP postulates that in Nature as many
phases as possible, which are allowed by the underlying theory,
should coexist. On the phase diagram of the theory it corresponds
to the special point -- the multiple point -- where many phases
meet. The vacuum energy densities of these different phases are
degenerate at the multiple point. In other words Nature should
adjust all the couplings of the SM (or any other model) such that
a number of degenerate vacua are realized. The MPP relations
between coupling constants can arise dynamically. For example a
mild form of locality breaking in quantum gravity, due to baby
universes \cite{3} say, is expected to precisely fine-tune the
couplings so that indeed several phases with degenerate vacua
coexist. Another possible origin for MPP could be a symmetry.
Supersymmetry is the best candidate for this role because all
global vacua in SUSY models are degenerate. Moreover the SUSY
scalar potential often contains a few flat directions with zero
vacuum energy.

The idea of the multiple point principle was applied to the pure
SM, by postulating that the Higgs effective potential has two
rings of minima in the Mexican hat with the same vacuum energy
density \cite{4} (the effective potential depends only on the
Higgs field norm and has two minima in it). The radius of the
little ring is at the electroweak vacuum expectation value of the
Higgs field, while the radius of the big one was assumed to be
near the Planck scale ($M_{Pl}\approx 10^{19}\,\mbox{GeV}$). These
two assumptions lead to rather precise predictions for the top
quark (pole) and Higgs boson masses \cite{4}
\be
M_t=173\pm 4\,\mbox{GeV}\, ,\qquad M_H=135\pm 9\, \mbox{GeV}\, .
\label{1}
\ee

In the present article MPP is used, in conjunction with supersymmetric
models, to deduce a size for the cosmological constant to be
compared with the value obtained
by astrophysical observations. We shall do this by assuming
{\em a priori} the existence of a supersymmetric phase in flat
(Minkowski) space, in addition to the phase in which we live.
Since the vacuum energy density
(cosmological constant) of supersymmetric states in flat Minkowski space
is just zero, the cosmological constant problem is thereby solved to
first approximation by assumption.
Now the strategy is to estimate the SUSY breaking energy contribution
in this most supersymmetric phase, which is the only contribution to
the cosmological constant in this case and, by virtue of MPP,
to assign the found value to all other phases and especially to the one
in which we live.

However such a procedure immediately raises the question as to why the most
supersymmetric phase is taken, among all the various ones, to get its
value for the cosmological constant transfered via MPP to all the other phases.
The suggested answer is that one should choose, for this purpose,
the phase with the smallest SUSY breaking and thus lowest cosmological
constant to be the decisive phase. It comes from the philosophy that
it is easier for MPP to tune some cancellation, in order to make a
quantity small, than it is to get its value strongly enhanced.

Of course we do not solve the cosmological constant problem
entirely in this work.
Indeed there is a murky point in the suggested procedure. In order to have
a tiny value of the cosmological constant in the phase where the
supersymmetry is broken severely, we must call upon supergravity (SUGRA).
In this case a hidden sector can give an
additional contribution to the total energy density cancelling ones from other
sources (like electroweak symmetry breaking in our phase for example).
At the same time, even in the vacuum where local supersymmetry
remains intact, the total energy density tends to be huge and negative.
It makes our initial assumption concerning the existence of a phase
with global SUSY in flat Minkowski space rather artificial.
An extra fine-tuning is required to obtain a viable solution of this type
and corresponds to searching for only a partial solution of
the cosmological constant problem. The aim of this paper is to
calculate the deviation from zero cosmological constant, once our
initial assumption is accepted.

This article is organized as follows: in the the next section
we describe the structure of
$(N=1)$ SUGRA models and discuss the mechanism of supersymmetry breaking;
we formulate our MPP supergravity model in section 3 and present
some numerical estimates of the vacuum energy
density in section 4. Our results are summarized in section 5.

\section{From supergravity to SM}

SUSY models clear the way to the unification of gauge interactions
with gravity. Such unification is carried out in the framework of
SUGRA models. Simplest $(N=1)$ supersymmetric models correspond to
$(N=1)$ supergravity. The full $(N=1)$ SUGRA Lagrangian \cite{7}
is specified in terms of an analytic gauge kinetic function
$f_a(\phi_{M})$ and a real gauge-invariant K$\Ddot{a}$hler function
$G(\phi_{M},\phi_{M}^{*})$, which depend on the chiral superfields
$\phi_M$. The function $f_{a}(\phi_M)$ determines the kinetic
terms for the fields in the vector supermultiplets and the gauge
coupling constants $Re f_a(\phi_M)=1/g_a^2$, where the index $a$
designates different gauge groups. The K$\Ddot{a}$hler function is a
combination of two functions
\be
G(\phi_{M},\phi_{M}^{*})=K(\phi_{M},\phi_{M}^{*})+
\ln|W(\phi_M)|^2\, , \label{2} \ee where
$K(\phi_{M},\phi_{M}^{*})$ is the K$\Ddot{a}$hler potential whose
second derivatives define the kinetic terms for the fields in the
chiral supermultiplets. $W(\phi_M)$ is the complete analytic
superpotential of the considered SUSY model. In this article
standard supergravity mass units are used:
$\ds\frac{M_{Pl}}{\sqrt{8\pi}}=1$.

Experimentally, of course,
SUSY can not be an exact symmetry at low energies and has to be broken
in such a way that quadratic divergences are not induced
(so-called soft SUSY breaking).
In SUGRA models local supersymmetry breaking happens in a hidden sector, which
contains singlet superfields $(h_m)$ under the SM
$SU(3)\times SU(2)\times U(1)$ gauge group.
These hidden superfields are introduced by hand in the simplest models .

In theoretically reliable SUGRA models
the form of the K$\Ddot{a}$hler function and the structure of the
hidden sector are fixed by an underlying renormalizable or even finite theory.
Nowadays the best candidate for the ultimate theory is $E_8\times E_8$
(ten dimensional)
heterotic superstring theory \cite{8}.
The strong coupling limit in this theory can be described by
eleven dimensional supergravity on a manifold with two ten--dimensional
boundaries (M--theory) \cite{9}.
Gauge multiplets of each $E_8$ gauge group are localized on a separate
boundary and
interact with multiplets of the other $E_8$ by virtue of gravitational forces.
Compactification of the extra dimensions on a Calabi-Yau
manifold leads to an effective supergravity and results
in the breaking of one $E_8$ to $E_6$ or its subgroups,
which play the role of gauge symmetries in the observable sector.
Multiplets of the remaining $E_8$
belong to the hidden part of the considered theory.
Although all hidden sector multiplets can give rise to violation
of supersymmetry, the minimal possible SUSY--breaking sector in
string models involves dilaton ($S$) and moduli ($T_m$) superfields.
The number of moduli varies from one string model to another.
But dilaton and moduli fields are always
present in four--dimensional heterotic superstrings,
because $S$ is related with the gravitational
sector while vacuum expectation values of $T_m$ determine the size
and shape of the compactified space.

After integration over hidden sector fields the superpotential of
SUGRA models generally looks like (see for example \cite{10})
\be
W(\phi_M)=W^{(tree)}(\phi_M)+W^{(ind)}(\phi_M)\, ,
\label{21}
\ee
where
\be
W^{(tree)}(\phi_M)=\frac{1}{6}Y'_{\alpha\beta\gamma}(h_m)C^{\alpha}
C^{\beta}C^{\gamma}+...\, \label{3} \ee is a classical
superpotential that depends on hidden $h_m$ (dilaton and moduli)
and observable $C^{\alpha}$ superfields. Generally supersymmetric
mass terms are assumed to be absent in the classical part of the
superpotential. They may be induced by non-perturbative
corrections that summarize the effects of integrating out the
hidden sector \cite{10}:
\be
W^{(ind)}(\phi_M)=\hat{W}(h_m)+\frac{1}{2}\mu'_{\alpha\beta}(h_m)C^{\alpha}
C^{\beta}+...\, .
\label{4}
\ee
Expanding the full K$\Ddot{a}$hler potential in powers of observable
fields $C^{\alpha}$,
we have \cite{10}-\cite{11}
\be
K=\hat{K}(h_m,h^{*}_m)+\tilde{K}_{\bar{\alpha}\beta}(h_m, h^{*}_m)
C^{*\bar{\alpha}}C^{\beta}+ \left[\frac{1}{2}Z_{\alpha\beta}(h_m,
h^{*}_m)C^{\alpha}C^{\beta}+h.c. \right]+...,
\label{5}
\ee
The dots in formulas (\ref{3})--(\ref{5}) stand for higher order terms
whose coefficients are suppressed by negative powers of $M_{Pl}$.
$\hat{W}(h_m)$ and $\hat{K}(h_m, h^{*}_m)$ are the superpotential and the
K$\Ddot{a}$hler potential of the hidden sector respectively. Notice
that the coefficients $Y'_{\alpha\beta\gamma},\,
\mu'_{\alpha\beta},\, \tilde{K}_{\bar{\alpha}\beta}$ and
$Z_{\alpha\beta}$ in the expansions (\ref{3})--(\ref{5}) depend on
the hidden sector fields in general. The bilinear terms associated
with $\mu'_{\alpha\beta}$ and $Z_{\alpha\beta}$ are often
forbidden by gauge invariance. However their appearance destroys the
$Z_3$ discrete symmetry that leads to the domain wall problem
\cite{12} and provides a viable solution for the so--called $\mu$
problem, in the context of the minimal supersymmetric standard
model (MSSM) \cite{13}.

The SUGRA scalar potential can be presented as a sum of $F$-- and D--terms
$V_{SUGRA}(\phi_M, \phi^{*}_M)=V_{F}(\phi_M, \phi^{*}_M)+V_{D}(\phi_M,
\phi^{*}_M)$, where the F--part is given by \cite{7},\cite{14}
\be
\ba{c}
V_{F}(\phi_M,\phi^{*}_M)=e^{G}\left(G_{M}G^{M\bar{N}}
G_{\bar{N}}-3\right)\, ,\\[3mm]
G_M \equiv\partial_{M} G\equiv\partial G/\partial \phi_M,
\qquad G_{\bar{M}}\equiv
\partial_{\bar{M}}G\equiv\partial G/ \partial \phi^{*}_M\, , \\[3mm]
G_{\bar{N}M}\equiv
\partial_{\bar{N}}\partial_{M}G=\partial_{\bar{N}}\partial_{M}K\equiv
K_{\bar{N}M}\, .
\ea
\label{6}
\ee
The matrix $G^{M\bar{N}}$ is the
inverse of the K$\Ddot{a}$hler metric $K_{\bar{N}M}$. If, at the
minimum of the scalar potential, hidden sector fields acquire vacuum
expectation values so that at least one of their auxiliary fields
\be
F^{M}=e^{G/2}G^{M\bar{P}}G_{\bar{P}}
\label{61}
\ee
is non-vanishing, then local SUSY is spontaneously broken. At the same
time a massless fermion with spin $1/2$ -- the goldstino, which is a
combination of the fermionic partners of the hidden sector fields
giving rise to the breaking of SUGRA, is swallowed by the gravitino
that becomes massive
$$
m_{3/2}=<e^{G/2}>\, .
$$
This phenomenon is called the super-Higgs effect \cite{15}.

Since the superfields of the hidden sector interact with the
observable ones only by means of gravity, they are decoupled from
the low energy theory. The only signal they produce are a set of
terms that break the global supersymmetry of the low-energy effective
Lagrangian of the observable sector in a soft way
\cite{16}--\cite{17}. The set of soft SUSY breaking parameters
includes: gaugino masses $M_a$, masses of scalar components of
observable superfields $m_{\alpha}$, trilinear
$A_{\alpha\beta\gamma}$ and bilinear $B_{\alpha\beta}$ scalar
couplings associated with Yukawa couplings and $\mu$--terms in the
superpotential of the considered SUSY model \cite{18}. Using the explicit
form of the SUGRA scalar potential (\ref{6}) and the expansion of
the K$\Ddot{a}$hler function in terms of observable superfields
(\ref{3})--(\ref{5}), one can compute soft SUSY breaking terms at
the Planck or Grand Unification scale. They are obtained by
substituting vacuum expectation values for the hidden sector fields
$h_m$ and corresponding auxiliary fields $F^{m}$, and taking the
flat limit \cite{19} where $M_{Pl}\to\infty$ but
$m_{3/2}$ is kept fixed. Then one is left with a global SUSY
Lagrangian plus the soft SUSY breaking terms listed above. All
non-renormalizable terms can be omitted, since they are suppressed
by inverse powers of $M_{Pl}$. Choosing the K$\Ddot{a}$hler metric
of the observable sector in the diagonal form
$\tilde{K}_{\bar{\alpha}\beta}=\tilde{K}_{\alpha}\delta_{\bar{\alpha}\beta}$
to avoid dangerous flavour--changing neutral current (FCNC)
transitions and assuming that, at the minimum of the SUGRA scalar
potential, the value of the cosmological constant equals zero
$<V(h_m)>=0$, one finds \cite{10}--\cite{11}
\be
\ba{c}
m^2_{\alpha}=m^2_{3/2}-F^{\bar{m}}F^{n}\partial_{\bar{m}}
\partial_{n}\ln\tilde{K_{\alpha}}\,,\\[3mm]
A_{\alpha\beta\gamma}=F^{m}\left[
\hat{K}_m+\partial_m\ln Y'_{\alpha\beta\gamma}-\partial_m
\ln(\tilde{K}_\alpha\tilde{K}_{\beta}\tilde{K}_{\gamma})\right]\,,\\[3mm]
M_a=\ds\frac{1}{2}(Ref_a)^{-1}F^{m}\partial_{m}f_a\, .
\ea
\label{7}
\ee
As usual the D-term contributions to SUSY breaking are neglected.
Explicit expressions
for the bilinear scalar couplings
$B_{\alpha\beta}$ are not given here, because they depend significantly on the
mechanism of $\mu$--term generation
(see \cite{11}).

The first term in the formula for $m_{\alpha}^2$ gives a
universal positive contribution to all soft scalar masses that, in
general, allows us to make scalar particles heavier than their
fermionic partners. The size of all soft SUSY breaking terms is
characterized by the gravitino mass scale. Therefore the gravitino
mass should not be very large, since the soft masses of the Higgs
bosons have to be of the order of the electroweak scale to ensure a
correct pattern for the $SU(2)\times U(1)$ symmetry breaking. A huge
mass hierarchy ($m_{3/2}\ll M_{Pl}$) can appear due to a
non-pertubative source of local supersymmetry breaking in the
hidden sector gauge group \cite{20}.

\section{Multiple Point Principle in SUGRA models}

Let us now consider SUGRA models which obey the multiple point
principle. This means that there must be two or even more
degenerate vacua in the considered models. In the Standard Model
and its renormalizable extensions the MPP conditions are attained
by adjusting arbitrary coupling constants. As mentioned in section
2, in SUGRA models there are two arbitrary functions that should
be fixed via MPP, in the same way as the coupling constants in
renormalizable theories, resulting in a set of degenerate vacua.

As described above a common paradigm implies that, at one minimum of
the scalar potential (\ref{6}), local supersymmetry is broken leading to the
appearance of soft terms in the effective Lagrangian of the observable
sector. Further we will treat this vacuum as the physical one, which
is realized in Nature, and denote the appropriate vacuum
expectation values of hidden fields as $h_m^{(1)}$.

However MPP inspired SUGRA models may have another minimum of the
scalar potential with the same energy density, where the
supersymmetry in the hidden sector is unbroken. Moreover we assume
that, in this second vacuum, the low energy limit of the
considered theory is described by a pure supersymmetric model in
flat Minkowski space. As discussed in the introduction, the last
requirement represents an extra fine--tuning because in general
the cosmological constant in SUGRA models is huge and negative. To
show this, let us suppose that, the K$\Ddot{a}$hler function has a
stationary point $\phi_M=\phi_M^0$, where $G_M=0$. Then it is easy
to check that this point is also an extremum of $V_{SUGRA}(\phi_M,
\phi_M^*)$. Since, according to Eq.(\ref{61}), the auxiliary
fields $F^M$ are ``proportional'' to the $G_M$, they vanish in its
vicinity and local supersymmetry remains intact. At the same time
the energy density is huge and negative. While all D--terms go to
zero near the extremum point of $G(\phi_M, \phi_M^*)$, the last
term in the brackets of Eq.(\ref{6}) for $V_F(\phi_M, \phi_M^*)$
gives a finite and negative contribution to the total density of
energy. Thus the cosmological constant in such SUGRA models is
less than or equal to $-3e^{\ds G(\phi_M^0, \phi_M^{0*})}$.

On the other hand, in flat Minkowski space the energy density of
any supersymmetric vacuum state is exactly zero.
The effective description of the second vacuum, in terms of a
supersymmetric one, is supposed to be valid down to very low
energies ($E\ll M_Z$) in the MPP inspired SUGRA model. Thus
all soft SUSY breaking terms induced into the observable sector
must vanish (with much higher accuracy than in the physical vacuum)
and particles from a single supermultiplet will have the same mass.
Since in the SUSY limit the graviton and gravitino are massless
in the flat space-time approximation, one obtains an additional
constraint on the value of the superpotential
of the hidden sector
\be
<\hat{W}(h_m^{(2)})>=0
\label{8}
\ee
where $h_m^{(2)}$ denote vacuum expectation values of the hidden
sector fields in the second
vacuum. Equation (\ref{8}) is nothing other than the extra fine-tuning
in our model that corresponds to giving up the complete
solution of the cosmological constant problem.

If condition (\ref{8}) is fulfilled then the last term in the
brackets of Eq.(\ref{6}), which led to the negative energy
density, vanishes. Taking into account that the K$\Ddot{a}$hler
metric of the hidden sector is positive definite, one can prove in
this case that the absolute minimum of the scalar potential
(\ref{6}) is achieved when
\be
\frac{\partial \hat{W}(h_m)}{\partial
h_k}\Biggl|_{h_m=h_m^{(2)}}=0\, .
\label{9}
\ee
Together with the superpotential of the hidden sector and its derivatives,
the energy density of the second vacuum and the auxiliary fields $F^M$
go to zero, verifying that supersymmetry really is unbroken.

In order to demonstrate how the conditions (\ref{8}) and
(\ref{9}) work, let us consider a particular example. For the sake
of simplicity, we restrict our consideration to the minimal SUGRA model
\cite{16}, \cite{19}, \cite{21} with K$\Ddot{a}$hler potential
\be
K(\phi_{M},\phi_{M}^{*})=\sum_m h_m h_m^{*}+\sum_{\alpha}
|C^{\alpha}|^2
\label{91}
\ee
which results in canonical kinetic
terms in the supergravity Lagrangian. A canonical choice for the
kinetic function $f_a(h_m)=const$ corresponds to $M_a=0$. Therefore
we assume a mild dependence of $f_a(h_m)$ on the hidden fields, so
that the gauge couplings in the physical and supersymmetric vacua do
not differ by much, i.e. $|f_a(h_m^{(1)})-f_a(h_m^{(2)})|\ll
f_a(h_m^{(1)})$.

Because the K$\Ddot{a}$hler metric $K_{\bar{N}M}$ and
its inverse are diagonal, an explicit form for the SUGRA scalar
potential of the hidden sector can be easily found:
\be
V_F^{hid}(h_m, h_m^{*})=e^{\ds\hat{K}(h_m,
h_m^*)}\Biggl(\sum_k\Biggl|\ds\frac{\partial
\hat{W}(h_m)}{\partial h_k} +h_k^*
\hat{W}(h_m)\Biggr|^2-3|\hat{W}(h_m)|^2\Biggr)\, .
\label{10}
\ee
Although in principle the potential (\ref{10}) takes positive
as well as negative values near the second minimum, where supersymmetry is
preserved, the energy density is always larger than or equal to zero:
\be
<V_F^{hid}(h_m^{(2)})>_{SUSY}=e^{\ds\hat{K}(h_m,
h_m^*)}\Biggl(\sum_k \Biggl|\ds\frac{\partial
\hat{W}(h_m)}{\partial
h_k}\Biggr|^2\Biggr)\Biggl|_{h_m=h_m^{(2)}}\, ,
\label{11}
\ee
while $h_m^{(2)}$ satisfy equations for extrema
\be
\sum_k\left(\ds\frac{\partial \hat{W}(h_m)}{\partial
h_k}\right)^*\Biggl[\ds\frac{\partial^2 \hat{W}(h_m)} {\partial
h_k\partial h_n}+h_n^*\frac{\partial\hat{W}(h_m)}{\partial
h_k}+h_k^*\frac{\partial \hat{W}(h_m)}{\partial h_n} \Biggr]=0\, .
\label{12}
\ee
In the minimization conditions (\ref{12}) we put
$\hat{W}(h_m)=0$. The index $n$ varies from $1$ to $N$, where $N$
is the number of scalar fields in the hidden sector which acquire
non--zero vacuum expectation values. From the equations that
determine the position of the stationary point of the SUGRA scalar
potential (\ref{12}) and the expression for the energy density in
the supersymmetric vacuum (\ref{11}), it becomes clear that the
deepest minimum is reached when the conditions (\ref{9}) are
satisfied and the value of the scalar potential (\ref{10}) equals
zero.

In the instance when the hidden sector contains only one singlet
superfield, the simplest superpotential that suits MPP is
\be
\hat{W}(S)=m_0(S+\beta)^2\, . \label{13} \ee If the parameter
$\beta=-\sqrt{3}+2\sqrt{2}$, the SUGRA scalar potential possesses
two degenerate minima with zero energy density at the classical
level. The appropriate hidden scalar potential and superpotential
as a function of the scalar component of the superfield $S$ are
shown in Figs. 1a and 1b. For large $|S|\gtrsim 1$ the SUGRA
potential grows rapidly because of the exponential factor $e^{\ds
|S|^2}$ that arises due to the first term in (\ref{91}). There are
three extremum points in the scalar potential. The left minimum
coincides with the stationary point of the superpotential
(\ref{13}), where it vanishes, so that supersymmetery is unbroken.
The right minimum is attained for $<S>=S_0=\sqrt{3}-\sqrt{2}$. In
this vacuum, the gravitino gets a mass $m_{3/2}=1.487\cdot m_0$
and the set of soft SUSY breaking terms is generated:
\be
m^2_{\alpha}=m^2_{3/2}\, , \qquad
A_{\alpha\beta\gamma}=(3-\sqrt{6})m_{3/2}\, . \label{14} \ee To
obtain these results, the explicit expressions for the
K$\Ddot{a}$hler potential (\ref{91}) and superpotential (\ref{13})
were substituted in formulas (\ref{2}), (\ref{61}) and (\ref{7}),
where the field $S$ was replaced by its vacuum expectation value
$S_0$. The predictions for the gaugino masses $M_a$ are not given
here, since we do not specify the dependence of the kinetic
function on the hidden field $S$.

More complex structure in the hidden sector superpotential can
lead to a scalar potential that has a few vacua in which the
supersymmetry of the full $N=1$ SUGRA Lagrangian is exact or only
spontaneously broken. The MPP requires the degeneracy of all the
vacua or at least the deepest physical and supersymmetric ones. If
the hidden sector involves more than one superfield, SUGRA models
may possess so--called vacuum valleys or flat directions. Then the
most preferable situation, from an MPP believer's point of view,
arises when many vacua or flat directions that might be
supersymmetric or not have the same energy density. However,
having one vacuum obeying the relations (\ref{8}) and (\ref{9})
means the existence of just one extra phase degenerate with our
own; this is only a beginning or a necessary condition for MPP in
the minimal SUGRA models. In general MPP means that there is a
number of degenerate minima $h_m^{(i)}$:
\be
\ba{c}
V(h_m^{(1)}, h_m^{(1)*})=V(h_m^{(i)},
h_m^{(i)*})\, , \\[2mm]
\ds\frac{\partial V(h_m, h^*_m)}{\partial h_k}\Biggl|_{h_m=h_m^{(1)}}=
\ds\frac{\partial V(h_m, h^*_m)}{\partial h_k}\Biggl|_{h_m=h_m^{(i)}}=0\, ,
\ea
\label{15}
\ee
Here $V(h_m, h_m^*)$ should be identified with the full SUGRA scalar potential
$V_{SUGRA}(h_m, h_m^*)$ or with its F-part $V_F(h_m, h^*_m)$ if all hidden
sector fields are singlets.

\section{The value of the cosmological constant}

In principle, the supersymmetry that remains intact in the second
vacuum can be broken dynamically at low energies (for recent
reviews see \cite{22}--\cite{23}). Indeed, even in the pure MSSM,
the beta function of the strong gauge coupling constant exhibits
asymptotically free behaviour ($b_3=-3$). Since in the minimal
SUGRA model the kinetic function does not depend on the hidden
superfields ($f_a(h_m)=const$), the values of the gauge couplings
at the unification scale and their running down to the scale
$M_{S}\simeq m_{3/2}$ are the same in both vacua. Below the scale
$M_S$ all superparticles in the physical vacuum decouple and the
corresponding beta function changes ($\tilde{b}_3=-7$). Using the
value of $\alpha^{(1)}_3(M_Z)\approx 0.118\pm 0.003$ \cite{24} and
matching condition $\alpha^{(2)}_3(M_S)=\alpha^{(1)}_3(M_S)$, one
finds the strong coupling in the second vacuum
\be
\ds\frac{1}{\alpha^{(2)}_3(M_S)}=\ds\frac{1}{\alpha^{(1)}_3(M_Z)}-
\frac{\tilde{b}_3}{4\pi}\ln\frac{M^2_{S}}{M_Z^2}\, .
\label{16}
\ee
Here $\alpha^{(1)}_3$ and $\alpha^{(2)}_3$ are the values of the strong
gauge couplings in the physical and second minima of the SUGRA scalar
potential.

At the scale $\Lambda_{SQCD}$, where the supersymmetric QCD interaction
becomes strong in the second vacuum
\be
\Lambda_{SQCD}=M_{S}\exp\left[{\frac{2\pi}{b_3\alpha_3^{(2)}(M_{S})}}\right]\,
\label{17}
\ee
the supersymmetry may be broken due to non--perturbative effects. If instantons
generate a repulsive superpotential \cite{22}, \cite{25} which lifts and
stabilizes the vacuum valleys in the scalar potential,
then a generalized O'Raifeartaigh mechanism can take
place inducing a non--zero value for the cosmological constant
\be
\Lambda \sim \Lambda_{SQCD}^4\, .
\label{18}
\ee

In Fig.2 the dependence of $\Lambda_{SQCD}$ on the SUSY breaking
scale $M_S$ is examined. Because $\tilde{b}_3 < b_3$ the QCD gauge
coupling below $M_S$ is larger in the physical minimum than in the
second one. Therefore the value of $\Lambda_{SQCD}$ is much lower
than in the Standard Model and diminishes with increasing $M_S$.
For the pure MSSM it varies from $10^{-25}M_{Pl}$ to
$10^{-30}M_{Pl}$, when $M_S$ grows from $100\,\mbox{GeV}$ to
$1000\,\mbox{TeV}$. From rough estimates of the energy density
(\ref{18}), it can be easily seen that
$\Lambda_{SQCD}=10^{-31}M_{Pl}$ gives the measured value of the
cosmological constant. If MSSM is supplemented by an additional
pair of $5+\bar{5}$ multiplets then $\Lambda_{SQCD}$ of the
required size can be reproduced even for $M_S=100\div 1000\,
\mbox{GeV}$.

Achieving the SUSY breaking at the scale $\Lambda_{SQCD}$ is actually not at
all easy. The discussion is different depending on whether the number of
flavours $N_f$ is larger or smaller than the number of colours $N_c$.
%Now, neglecting Yukawa couplings, a non-perturbative superpotential
%of repulsive form
%indeed emerges in supersymmetric QCD, if the number of flavours
%($N_f$) is less than the number of colours ($N_c$).
%However this condition is not satisfied
In the MSSM and its simplest extensions where $N_c=3$ and $N_f=6$
%In the latter case,
the generated superpotential has
a polynomial form \cite{23}, \cite{26}. The absolute minimum of the SUSY
scalar potential is then achieved when all the superfields, including
their F- and D-terms, acquire zero vacuum expectation
values preserving supersymmetry. This result throws some doubt on
our scenario for a tiny cosmological constant, which is based on
Eq. (\ref{18}).

Another method of breaking SUSY is by the appearance of gaugino condensation
$\bar{\lambda}_{a}\lambda_{a}$. The gaugino condensation itself does not
lead to the spontaneous breakdown of global supersymmetry \cite{27}.
But if a non--trivial dependence of the gauge kinetic function on the
hidden sector fields is assumed then the corresponding auxiliary fields
\be
F^{i}=e^{G/2}G^{i \bar{j}}G_{\bar{j}}-\frac{1}{4}
G^{i\bar{j}}\frac{\partial f(h_m)}{\partial h_j}\bar{\lambda}_a\lambda_a+...
\label{19}
\ee
get an extra contribution which is proportional to
$<\bar{\lambda}_a\lambda_a>\simeq \Lambda_{SQCD}^3$ resulting in local
supersymmetry breaking \cite{20} and a non--zero vacuum energy density
\be
\Lambda \sim \frac{\Lambda_{SQCD}^6}{M_{Pl}^2}\, .
\label{20}
\ee
Unfortunately the gaugino condensation is not likely to occur if $N_f>N_c$.

However the above disappointing facts concerning dynamical SUSY breaking
were revealed in the framework of pure supersymmetric QCD,
where all Yukawa couplings were supposed to be
small or even absent. At the same time the t--quark Yukawa coupling
in the MSSM is of the same order of magnitude as the strong gauge
coupling at the electroweak scale. Therefore it can change the above results
drastically. We plan to continue our investigations of supersymmetry
breaking in the pure MSSM and its extensions.

\section{Conclusions}

In the present article we have applied the multiple point
principle assumption to $(N=1)$ supergravity. At first we reviewed
the structure of the $(N=1)$ SUGRA Lagrangian and local
supersymmetry breaking via the hidden sector. Explicit expressions
for the soft SUSY breaking parameters in terms of the
K$\Ddot{a}$hler and gauge kinetic functions were also collected.
The MPP inspired SUGRA model we considered implies that the
corresponding scalar potential contains at least two degenerate
minima. In one of them local supersymmetry is broken in the hidden
sector at the high energy scale $(\sim 10^{10}-10^{12}\,
\mbox{GeV})$, inducing a set of soft SUSY breaking terms for the
observable fields. In the other vacuum the low energy limit of the
considered theory is described by a pure supersymmetric model in
flat Minkowski space. This second minimum is realized if the
superpotential of the hidden sector has an extremum point where it
goes to zero. The stationary point of the superpotential coincides
with the position of the second minimum of the SUGRA scalar
potential. The energy density and all auxiliary fields $F^M$ of
the hidden sector vanish in the second vacuum preserving
supersymmetry. The simplest SUGRA model, where the MPP conditions
are satisfied, has been discussed and the predictions for the soft
masses and trilinear scalar couplings have been obtained.

Non--perturbative effects in the observable sector can give rise to
supersymmetry breakdown in the second vacuum (phase). In this case
the value of the energy density is determined
by the scale where the gauge interactions become strong. Numerical
estimates have been carried out in the framework of the pure MSSM.
They reveal that the corresponding scale is naturally low
($\Lambda_{SQCD}\simeq 10^{-30}-10^{-25}\,M_{Pl}$) providing a tiny
energy density of the second phase. The crucial idea is then to use
MPP to transfer the energy density or cosmological
constant from this second vacuum into all other vacua, especially into
the physical one in which we live. In such a way we have suggested
an explanation of why the observed value of the cosmological
constant has the tiny value it has.

The trouble with the considered approach is that the dynamical breakdown of
supersymmetry looks rather questionable, in models which involve
QCD with more flavours than colours (as in the SM and its simplest
SUSY extensions). However the strong interaction between the
Higgs and t-quark superfields in the superpotential, which has always
been ignored in previous considerations, could play a decisive role.
We intend to study this problem in more detail.

\vspace{10mm}
\noindent
{\Large \bf Acknowledgements}
\vspace{2mm}

\noindent
The authors are grateful to S.Bludman, D.Kazakov, O.Kancheli, V.Novikov,
L.Okun, M.Shifman, K.A.Ter-Martirosyan, M.Vysotsky and P.Zerwas for
fruitful discussions and A.Anisimov and M.Trusov for useful comments.
R.Nevzorov and H.B.Nielsen are indebted to the DESY Theory Group for
hospitality extended to them in 2002 when this project was started.
R.Nevzorov is also grateful to Alfred Toepfer Stiftung for the
scholarship and for the favour disposition
during his stay in Hamburg (2001--2002). C.Froggatt and H.B.Nielsen
would like to thank ITEP for inviting them to the XXXI-st ITEP Winter
School where this paper was completed and L.Bergamin for interesting
discussions.

\vspace{2mm}
\noindent
This work was supported by the Russian Foundation for Basic Research (RFBR),
projects 00-15-96786, 00-15-96562 and 02-02-17379.

\newpage
\noindent
{\Large \bf Figure captions}
\vspace{5mm}

\noindent {\bf Fig. 1.}\, (a) the scalar potential and (b) the
superpotential for the simplest SUGRA model where the MPP
conditions are satisfied. The standard supergravity mass units are
used.

\vspace{5mm} \noindent {\bf Fig. 2.}\, The value of
$\log\left[\Lambda_{SQCD}/M_{Pl}\right]$ versus $\log M_S$. The
thin and thick solid lines correspond to the pure MSSM and the
MSSM with an additional pair of $5+\bar{5}$ multiplets
respectively. The dashed and dash--dotted lines represent the
uncertainty in $\alpha_3(M_Z)$. The upper dashed and dash-dotted
lines are obtained for $\alpha_3(M_Z)=0.124$, while the lower ones
correspond to $\alpha_3(M_Z)=0.112$. The horizontal line
represents the measured value of $\Lambda^{1/4}$. The SUSY
breaking scale is given in GeV.

\newpage
\begin{center}
{\Large $\frac{V(S)}{m_0^2}$}\\ \vspace{2mm}
{\hspace*{-20mm}\includegraphics[height=100mm,keepaspectratio=true]{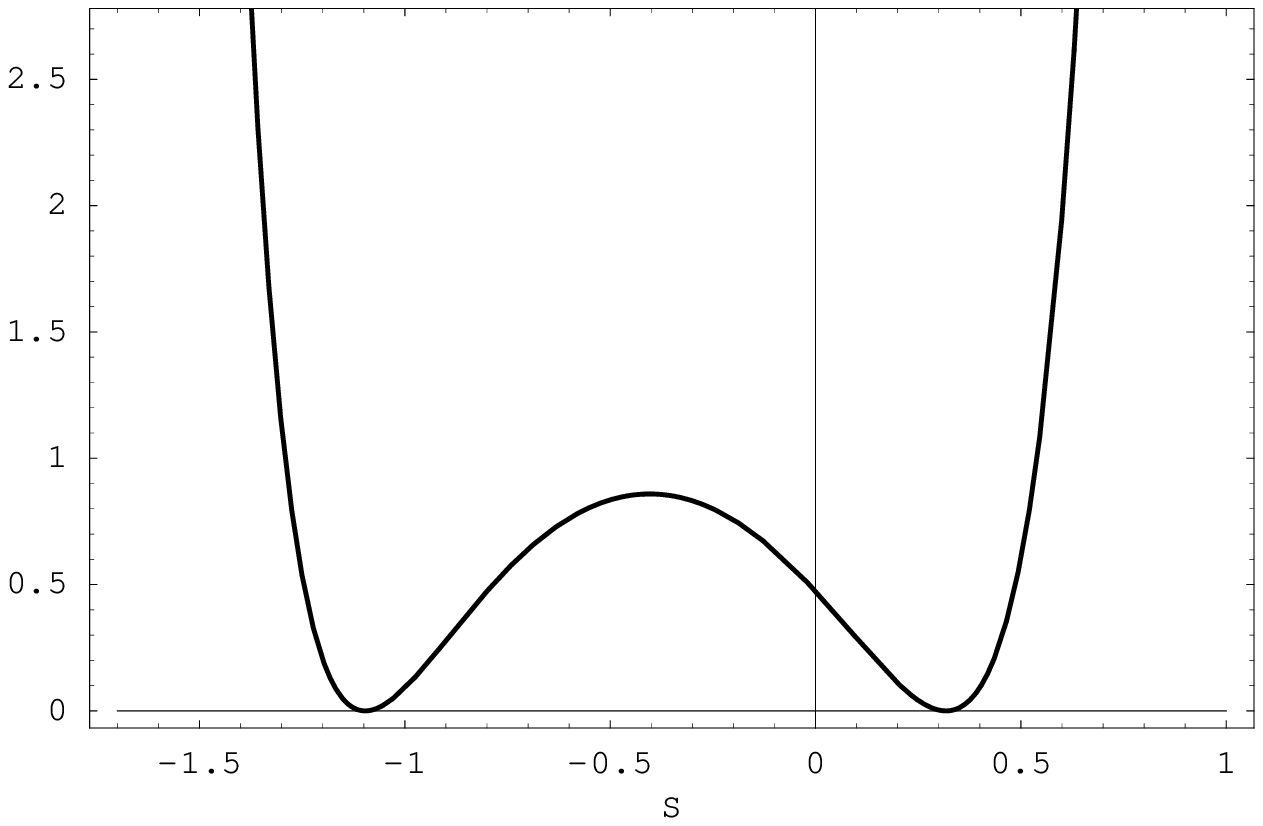}}\\
{\large\bfseries Fig.1a}\\ \vspace{0.5cm} {\Large
$\frac{W(S)}{m_0}$}\\ \vspace{2mm}
{\hspace*{-20mm}\includegraphics[height=100mm,keepaspectratio=true]{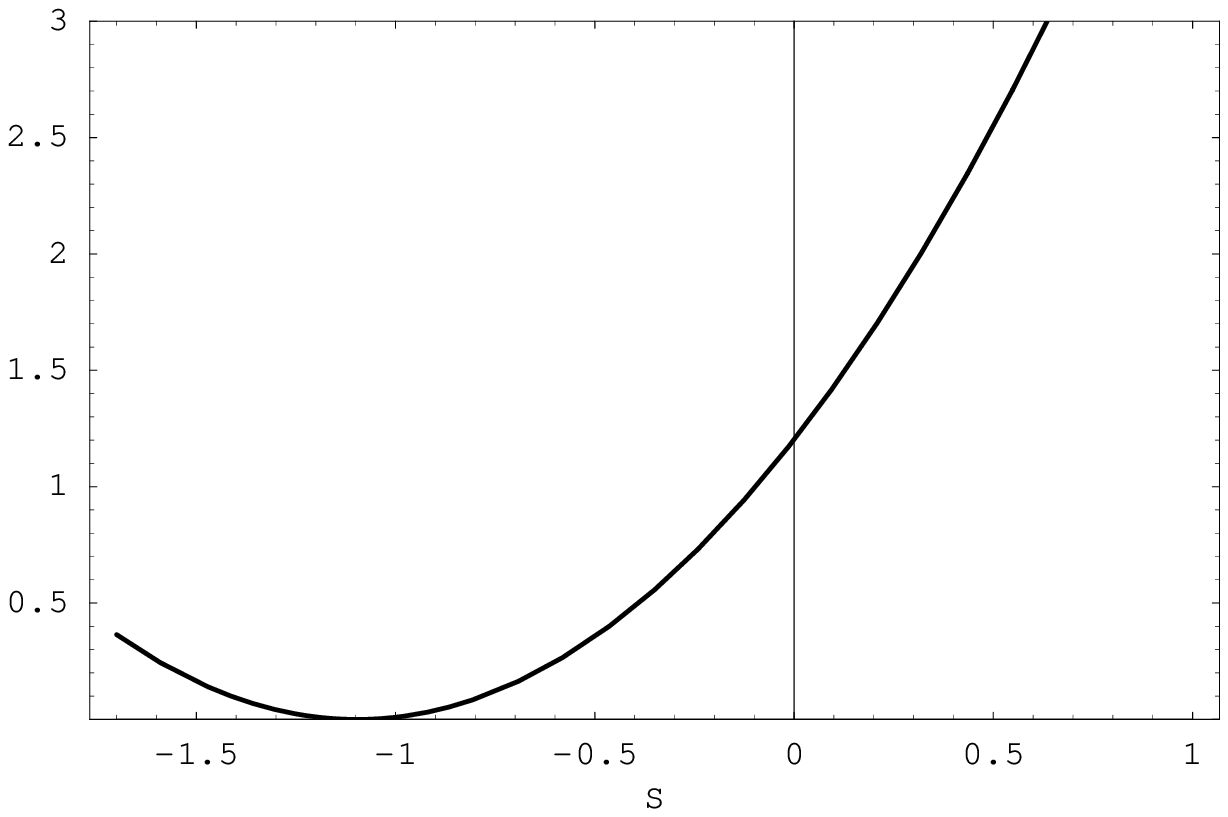}}\\
{\large\bfseries Fig.1b}
\end{center}

\newpage
\begin{center}
{\Large $\log[\Lambda_{SQCD}/M_{Pl}]$}\\ \vspace{2mm}
{\hspace*{-20mm}\includegraphics[height=100mm,keepaspectratio=true]{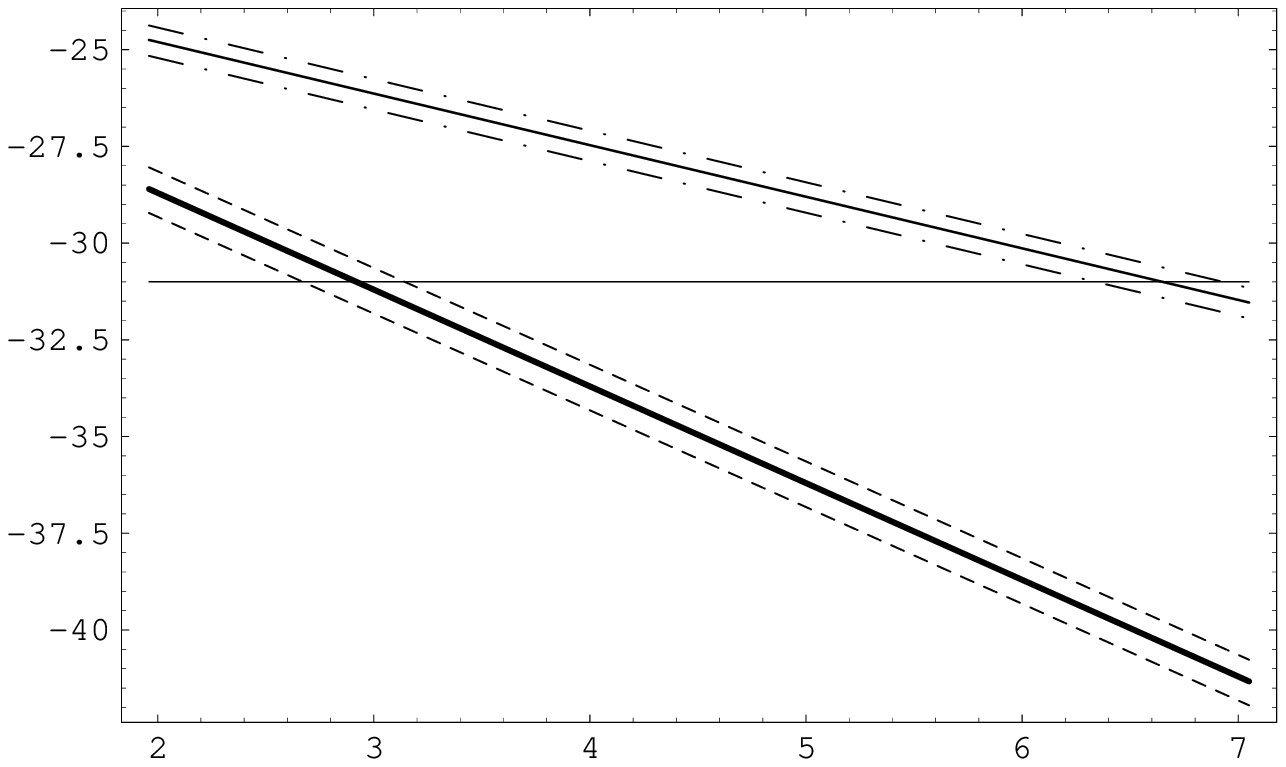}}\\
{\Large $\log[M_S]$}\\[3mm] {\large\bfseries Fig.2}
\end{center}


\newpage
\begin{thebibliography}{99}

\bibitem{0} A.G.Riess {\itshape et al.}, Astron.J. {\bf 116}, 1009 (1998);
S.Perlmutter {\itshape et al.}, Astrophys.J. {\bf 517}, 565 (1999);
C.Bennett {\itshape et al.}, astro-ph/0302207;
D.Spergel {\itshape et al.}, astro-ph/0302209.
\bibitem{2} D.L.Bennett, H.B.Nielsen, Int.J.Mod.Phys. A {\bf 9}, 5155 (1994);
{\it ibid} {\bf 14}, 3313 (1999);
D.L.Bennett, C.D.Froggatt, H.B.Nielsen, in {\itshape Proceedings of the
27th International Conference on
High energy Physics, Glasgow, Scotland, 1994}, p.557;
{\itshape Perspectives in Particle Physics '94, World Scientific, 1995},
p. 255, ed. D. Klabu\u{c}ar, I. Picek and D. Tadi\'{c}
[arXiv:hep-ph/9504294].
\bibitem{3} S.W.Hawking, Phys.Rev. D {\bf 37}, 904 (1988);
S.Coleman, Nucl.Phys. B {\bf 307}, 867 (1988);
{\it ibid} B {\bf 310}, 643 (1988);
T.Banks, Nucl.Phys. B {\bf 309}, 493 (1988).
\bibitem{4} C.D.Froggatt, H.B.Nielsen, Phys.Lett. B {\bf 368}, 96 (1996).
\bibitem{7} H.P.Nilles, Phys.Rep. {\bf 110}, 1 (1984);
A.B.Lahanas, D.V.Nanopoulos, Phys.Rep. {\bf 145}, 1 (1987).
\bibitem{8} M.B.Green, J.H.Schwarz, E.Witten, {\itshape Superstring Theory, Cambridge Univ. Press, Cambridge, 1987}.
\bibitem{9} P.Horava, E.Witten, Nucl.Phys. B {\bf 460}, 506 (1996); Nucl.Phys. B {\bf 475}, 94 (1996).
\bibitem{10} V.S.Kaplunovsky, J.Louis, Phys.Lett. B {\bf 306}, 269 (1993).
\bibitem{11} A.Brignole, L.E.Iba$\tilde{n}$ez, C.Mu$\tilde{n}$oz, Nucl.Phys. B {\bf 422}, 125 (1994),
Erratum: B {\bf 436}, 747 (1995); C.Mu$\tilde{n}$oz, hep-th/9507108;
A.Brignole, L.E.Iba$\tilde{n}$ez, C.Mu$\tilde{n}$oz, hep-ph/9707209.
\bibitem{12} Ya.B.Zel'dovich, I.Yu.Kobzarev, L.B.Okun, Sov.Phys.JETP {\bf 40}, 1 (1975);
S.A.Abel, S.Sarkar, P.L.White, Nucl.Phys. B {\bf 454}, 663 (1995).
\bibitem{13} J.E.Kim, H.P.Nilles, Phys.Lett. B {\bf 138}, 150 (1984), Phys.Lett. B {\bf 263}, 79 (1991);
G.F.Giudice, A.Masiero, Phys.Lett. B {\bf 206}, 480 (1988);
E.J.Chun, J.E.Kim, H.P.Nilles, Nucl.Phys. B {\bf 370}, 105 (1992);
J.A.Casas, C.Mu$\tilde{n}$oz, Phys.Lett. B {\bf 306}, 288 (1993).
\bibitem{14} E.Gremmer, S.Ferrara, L.Girardello, A.Van Proeyen, Phys.Lett. B {\bf 116}, 231 (1982),
Nucl.Phys. B {\bf 212}, 413 (1983).
\bibitem{15} S.Deser, B.Zumino. Phys.Rev.Lett. {\bf 38}, 1433 (1977);
E.Gremmer, B.Julia,J.Scherk, P.van Nieuwenhuizen, S.Ferrara, L.Girardello,
Phys.Lett. B {\bf 79}, 231 (1978), Nucl.Phys. B {\bf 147}, 105 (1979).
\bibitem{16} H.P.Nilles, M.Srednicki, D.Wyler, Phys.Lett. B {\bf 120}, 345 (1983).
\bibitem{17} L.Hall, J.Lykken, S.Weinberg, Phys.Rev. D {\bf 27}, 2359 (1983);
S.K.Soni, H.A.Weldon, Phys.Lett. B {\bf 126}, 215 (1983).
\bibitem{18} L.Girardello, M.T.Grisaru, Nucl.Phys. B {\bf 194}, 65 (1982).
\bibitem{19} R.Barbieri, S.Ferrara, C.Savoy, Phys.Lett. B {\bf 119}, 343 (1982).
\bibitem{20} H.P.Nilles, Int.J.Mod.Phys. A {\bf 5}, 4199 (1990).
\bibitem{21} A.H.Chamseddine, R.Arnowitt, P.Nath, Phys.Rev.Lett. {\bf 49}, 970 (1982).
\bibitem{22} M.Shifman, A.Vainshtein, hep-th/9902018.
\bibitem{23} D.S.Gorbunov, S.L.Dubovskii, S.V.Troitski, Usp.Fiz.Nauk. {\bf 169}, 705 (1999).
\bibitem{24} Review of Particle Properties, Eur.Phys.J. C {\bf 3}, 1 (1998).
\bibitem{25} I.Affleck, M.Dine, N.Seiberg, Nucl.Phys. B {\bf 241}, 493 (1984);
I.Affleck, M.Dine, N.Seiberg, Nucl.Phys. B {\bf 256}, 557 (1985);
V.Novikov {\itshape et al.}, Nucl.Phys. B {\bf 260}, 157 (1985).
\bibitem{26} N.Seiberg, Nucl.Phys. B {\bf 435}, 129 (1995).
\bibitem{27} G.Veneziano, S.Yankielowicz, Phys.Lett. B {\bf 113}, 231 (1982).


\end{thebibliography}
\end{document}